# Uniaxial anisotropy and low-temperature antiferromagnetism of $Mn_2BO_4$ single crystal

N.V. Kazak[1], M.S. Platunov[1], Yu.V. Knyazev[2], N.B. Ivanova[2], O.A. Bayukov[1], A.D. Vasiliev[1,2], L.N. Bezmaternykh[1], V.I. Nizhankovskii[4], S.Yu. Gavrilkin[5], K.V. Lamonova[6], and S.G. Ovchinnikov[1,2,3]

[1]*L.V. Kirensky Institute of Physics, SB of RAS, 660036 Krasnoyarsk, Russia*
[2]*Siberian Federal University, 660074 Krasnoyarsk, Russia*
[3]*Siberian State Aerospace University, 660014 Krasnoyarsk, Russia*
[4]*International Laboratory of High Magnetic Fields and Low Temperatures, PL-53421 Wroclaw, Poland*
[5]*P.N. Lebedev Physical Institute of RAS, 119991 Moscow, Russia*
[6]*O.O. Galkin Donetsk Institute for Physics and Engineering, National Academy of Sciences of Ukraine, 83114 Donetsk, Ukraine*

**Corresponding author:**
Dr. Natalia Kazak
Laboratory of Physics of Magnetic Phenomena
Kirensky Institute of Physics, Siberian Branch of Russian Academy of Science (IP SB RAS)
Academgorodok 50/38, Krasnoyarsk, 660036, Russia
+7(3912)49-45-56
nat@iph.krasn.ru

**Abstract** The $Mn_2BO_4$ single crystals have been grown using the flux technique. The careful study crystal structure and magnetic properties have been carried out. The antiferromagnet transition at $T_N = 26$ K has been traced through the *dc* magnetization and specific heat temperature dependences. The magnetic uniaxial anisotropy has been detected with easy axis of magnetization lying in *ab*-plane. A reduction of the effective magnetic moment value is assigned to the non-quenched orbital moment of Jahn-Teller $Mn^{3+}$ ions. Based on the superexchange interactions calculations the magnetic behavior is discussed.

**Keywords:** antiferromagnets, oxides, magnetic anisotropy, exchange interactions

**PACS:** 75.50.Ee, 75.30.Gw, 78.70.Dm, 75.30.Et

## 1. Introduction

It is well known that low dimensionality in the crystal and magnetic structure plays an important role in the physics of magnetic crystals. The experimental and theoretical investigations have revealed properties of the materials with low dimensionality to be quite different from their bulk analogues. This difference expresses, first of all, in the rich variety of phases and phase transitions caused by high degree of degeneration and extraordinary sensitivity to external influences. From this point of view the borates of transition and rare-earth metals with quasi-low dimension crystal structure and unique magnetic and optic properties are the perspective objects for the fundamental and practical investigations. These materials form a wide class of narrow-band oxide semiconductors intensely researched during the last years [1-5].

The oxyborates $M^{2+}M^{3+}BO_4$ with the warwickite structure attract attention due to a wide row of isomorphic substitutions. There is a large variety of natural and synthetic warwickites containing rare-earth, alkaline earth and transition metal ions ($M^{2+}$ = Mg, Co, Mn, Fe, Ni, … and $M^{3+}$ = Ti, V, Cr, Fe, Mn, In, Lu, Yb, Tm, …) [6-8]. The metal ions occupy two structurally distinct octahedral sites usually labeled as 1 and 2. Four octahedra linked by sharing edges to form the *row* of 2-1-1-2. The *rows* are connected by sharing edges forming quasi low-dimensional *ribbons* extending along crystalline *c*-axis. The heterometallic warwickites ($M^{2+} \neq M^{3+}$) are naturally disordered materials since each metal crystalline site may be occupied by any one of the two metals. Most hetero-metallic warwickites show typical spin-glass transition [9, 10].

Only two homometallic warwickites ($M^{2+} = M^{3+}$) are available now: $Fe_2BO_4$ and $Mn_2BO_4$ [11, 12]. Both compounds display the charge ordering (CO). The nature of CO in the warwickites is the question of the hot discussion. The temperature dependence of CO in $Fe_2BO_4$ was extensively investigated by resistivity and differential scanning calorimetry measurements, Mössbauer spectroscopy, synchrotron x-ray scattering, transmission electron microscopy, and electronic-structure calculations [13-16]. It is supposed that the low-temperature phase ($T_{CCO}$< 280 K) is a commensurately charge ordered with integer iron valence separation $Fe^{2+}$ and $Fe^{3+}$ alternating in the *a*-axis direction. The intermediate temperature range ($T_{CCO}$< $T$ < $T_{CO}$, where $T_{CO}$ =340 K) is characterized by the onset of the temperature-dependent lattice incommensurate CO accompanying by coexisting mobile and immobile carriers. There is a valence fluctuating state ($Fe^{2+}$ - $Fe^{3+}$ electron hopping) in the high-temperature interval ($T$>$T_{CO}$) where the structural transformation from orthorhombic → monoclinic symmetry takes place. In $Mn_2BO_4$ the CO appears to be related to strong Jahn-Teller distortion of Mn1O$_6$ octahedra. The CO of the kind Mn$^{2+}$(2)-Mn$^{3+}$(1)-Mn$^{3+}$(1)-Mn$^{2+}$(2) and relevant orbital ordering ($d_z^2$) occurs. From the magnetic point of view the Fe$_2$BO$_4$ was found to be a *L*-type ferrimagnet with transition temperature of $T_N$ = 155 K [17]. As for the Mn$_2$BO$_4$, the situation is more intriguing. Nowadays this compound was synthesized in three forms: single crystals [12], powder polycrystals [18-20], and necklace-like nanofibres [21]. The magnetic characterization was done for two first forms Mn$_2$BO$_4$ and the results obtained are dramatically controversial. On one hand, the magnetization, the specific heat, and ESR studies of polycrystalline samples have been revealed the antiferromagnetic transition at $T_N$ = 104 K and weak ferromagnetism below 70 K. Recent neutron diffraction have shown the occurring the long-range antiferromagnet order only below 26 K, while the ferrimagnetic transition at 42 K was found from the magnetization measurements. From other hand, the weak anomaly at ~25 K was observed from the magnetic susceptibility measurements performed on the single crystals. The reason of this dramatic disagreement probably lies in the samples quality. The authors of [18-20] marked the presence of magnetic impurities in the form of Mn$_2$O$_3$ and Mn$_3$O$_4$ oxides.

Thus, there is no clear understanding of either type of magnetic order or the temperature of magnetic phase transition in Mn$_2$BO$_4$. This paper reports first magnetization and specific heat measurements carried out on the Mn$_2$BO$_4$ single crystals. The experimental observation of magnetic phase transition at $T_N$ = 26 K strongly supports for the neutron diffraction results. The long-range antiferromagnetic order occurs below $T_N$. The magnetic measurements performed for two directions of the applied field relative the crystalline *c*-axis allowed to reveal the uniaxial anisotropy. Based on the superexchange interactions analysis the magnetic behavior is explained and possible spin configuration for the ordered state of Mn$_2$BO$_4$ is offered.

2. **Experimental procedure**

The solid state reaction method was found to give rise the difficulties in the preparing of pure samples [18-20]. The main synthesis problem relates to the frequent existence of concurring phases with different crystal structures. This problem have been discussed in detail in the well known systematic work by Capponi [7], where it has shown that attempts to grow oxyborates Co$^{2+}$Ga$^{3+}$BO$_4$, Co$^{2+}$Cr$^{3+}$BO$_4$, Co$^{2+}$Sc$^{3+}$BO$_4$ with the warwickite structure have leaded to concomitant ludwigite phases Co$^{2+}_2$Ga$^{3+}$BO$_5$, Co$^{2+}_2$Cr$^{3+}$BO$_5$, Co$^{2+}_2$Sc$^{3+}$BO$_5$. Due to this reason the exact parameters of growing process are of great importance for the successful preparation of single-phase material. At present work these parameters were defined after some probe reactions and X-ray diffraction control.

The solution has been made by the step by step melting of $B_2O_3$ oxide (6.9 g), $Bi_2Mo_3O_{12}$ (51 g), $Mn_2O_3$ (17.7 g) and $Na_2CO_3$ (4.1 g) at $T_1=1100^0$ C during 3 hours. Then the temperature was rapidly lowered to $T_2=970^0$ C followed by a slow cooling at a rate of 4° C a day. In two days the crucible was pulled out from the furnace and the solution was removed. Single crystals spontaneously formed on the walls of the crucible were rinsed with aqueous nitric acid at room temperature. The crystals were in the form of black needles up to 12 mm long, and the cross sectional area was smaller than 1.0× 0.5 mm.

The room temperature X-ray diffraction measurements were carried out using X-ray diffractometer SMART APEX II (MoKα radiation, CCD detector).

The field magnetization was measured using the handmade vibrating samples magnetometer (VSM) at the International Laboratory of High Magnetic Fields and Low Temperatures (Wroclaw, Poland). The *dc* magnetization has been measured as function of the temperature and applied magnetic field up to 140 kOe. The temperature interval was 1.8 - 300 K. The measurements were carried out for two directions of the external magnetic field relative to the crystallographic *c*-axis, which coincides with the needle's axis. The holder underground contribution was subtracted from the integral signal and the corrections associated with form anisotropy were taken into account.

The specific heat measurements have been done by the relaxation technique on commercial PPMS Quantum Design platform in the entire temperature interval ($T = 2–300$ K). The experimental error didn't exceed 1% for all temperatures.

### 3. Experimental results

*3.1. X-ray diffraction and normal coordinates calculation*

In this section, we present some crystallographic data, which are relevant for theoretical discussion below. The $Mn_2BO_4$ has a monoclinic unit cell (*P2₁/n* space group), with the angle *β* ≈ 90.751° slightly different from 90° (Table 1). No impurity phases have been detected by means of X-ray diffraction. All parameters are in good agreement with those reported earlier [12, 20]. The metal ions have two distinct positions labeled as 1 and 2, which are at general 4*e* Wyckoff position, oxygen atom has four distinct positions and boron have only one position. The atomic coordinates, isotropic displacement parameters, selected bond lengths and angles are listed in the Supplemental Materials (SM) Tables SM1 - SM3 [22].

The $Mn1O_6$ octahedron is considerably smaller then that the $Mn2O_6$ one as deduced from the average <M-O> bond length (2.065 instead 2.210 Å). The smaller <M - O> distances lead to increasing oxidation state and can indicate that the $Mn^{3+}$ ions prefer the $M1O_6$ octahedra, while $Mn^{2+}$ occupies the M2$O_6$ ones. As it was expected the shortest distances (less than 1.5 Å) are B - O inside the $BO_3$ triangle, which is the most tightly bound group in oxyborate structures.

Table 1. Crystal data and structure refinement of $Mn_2BO_4$.

| Empirical formula | $Mn_2BO_4$ |
|---|---|
| Formula weight (g mol$^{-1}$) | 184.69 |
| Crystal system | monoclinic |
| Space group | $P2_1/n$ |
| Unit cell parameters (Å, deg) | |
| $a$ | 9.2934(5) |
| $b$ | 9.5413(5) |
| $c$ | 3.2475(2) |
| $\beta$ | 90.7510(10) |
| Unit cell volume (Å$^3$) | 287.93(3) |
| Z | 4 |
| Calculated density (g cm$^3$) | 4.26023 |
| Radiation | MoK$\alpha$ |
| Wavelenght, $\lambda$ (Å) | 0.71073 |
| Temperature (K) | 296 |
| Crystal shape | Needle (along $c$) |
| Abs. coefficient (mm$^{-1}$) | 8.581 |
| F(000) | 348 |
| $\Theta$ range (deg) | 3.06 - 34.00 |
| Limiting indices | $-14 \leq h \leq 14$ |
| | $-14 \leq k \leq 14$ |
| | $-5 \leq l \leq 4$ |
| Reflections collected | 4701 |
| Reflections independent | 1157 |
| Data / restraints / parameters | 1157 / 0 / 65 |
| Extinction coefficient | 0.200(4) |
| GooF | 1.173 |
| Final $R$ indices | |
| $R$1 | 0.0184 |
| $wR$2 | 0.0422 |
| $R$ indeces (all data) | |
| $R$1 | 0.0199 |
| $wR$2 | 0.0428 |

The $Mn^{2+}$ and $Mn^{3+}$ distribution over the metallic sites can be studied by means of bond valences sums (BVS) calculation [23]. These empirical estimations predict atomic charges of 3.20/2.95 for Mn1 and 2.01/1.85 for Mn2, when bond valence parameters are related to $Mn^{2+}/Mn^{3+}$. So, there is a clear propensity of $Mn^{3+}$ to occupy the site 1 with atomic charge 2.95, while for $Mn^{2+}$ it is the site 2 (atomic charge 2.01). The obtained values are in agreement with the BVS calculation results reported by Norrestam [12] confirming that $Mn_2BO_4$ is a transition metal oxyborate with the explicit charge ordering. The BVS value for the boron atom was found to be 2.96 close to the formal valence 3+.

Both Mn1$O_6$ and Mn2$O_6$ octahedra are distorted. The distortions of the coordination octahedra can be described by the normal coordinates $Q_\alpha$ ($\alpha$= 1, 2, … 3$N$-3; $N$ – number of ligands), which are linear combinations of the Cartesian coordinates of oxygen, and classified according to the irreducible representations of the coordination complex symmetry (Tables 2), in terms of the $O_h$ symmetry group (Table SM4 [22]). The $Q_1$ coordinate describes high-symmetry distortions like the breathing-mode. The other normal coordinates correspond to low – symmetry distortions like JT ($Q_2$ and $Q_3$) and trigonal ($Q_4$, $Q_5$, $Q_6$) ones. The $Q_3$ coordinate presents tetragonal octahedral distortion along the $z$ – axis, whereas the $Q_2$ ones corresponds to the distortions with rhombic symmetry. The observed JT distortion is a combination of the normal modes $Q_1$, $Q_2$, $Q_3$. There are a lager axial elongation of the Mn1O$_6$ octahedron along the O1-Mn1-O3 axis with the average axial radii of 2.325 Å and the compression of the other four Mn1-O bonds (the average planar radii is 1.935 Å), which suggests a $d_z^2$ orbital ordering at the Mn1 site as will be discussed below. Contrary, the bonds distribution in Mn2O$_6$ is so that two long bonds (O1, O4) and two medium bonds (O2, O3) are roughly coplanar with an average radius of 2.238 Å, while two remain short bonds (O2, O4) are axial with the average radii 2.155 Å (see Fig. SM1 and Table SM3 [22]). It can be noted that both normal coordinates $Q_2, Q_3$ are considerably prevailing for M1 site, testifying the pronounced JT distortion of Mn1O$_6$ octahedron. The trigonal distortions are comparable for both types of the octahedra.

**Table 2.** The normal coordinates and ligand's displacement (Å) for metal ions in the M1O$_6$ and M2O$_6$ octahedral complexes. The $R_0$ is the M - O distance in the undistorted octahedron, that are accepted such in order to provide a zero value of $Q_1$. $R_0$ is 2.054 and 2.092 Å, for Mn1 and Mn2 respectively.

| Normal coordinates | Mn1 | | Mn2 | |
|---|---|---|---|---|
| | normal coordinate | displacement | normal coordinate | displacement |
| $Q_2$ | -0.3937 | -0.1968 | 0.0330 | 0.0165 |
| $Q_3$ | -0.2122 | -0.0919 | 0.0828 | 0.0358 |
| $Q_4$ | -0.3347 | -0.1674 | -0.2095 | -0.1048 |
| $Q_5$ | -0.2045 | -0.1023 | -0.4691 | -0.2345 |
| $Q_6$ | -0.2612 | -0.1306 | -0.2403 | -0.1201 |
| $Q_7$ | 0.0828 | 0.0414 | -0.0703 | -0.0351 |
| $Q_8$ | 0.1464 | 0.0732 | -0.4739 | -0.2369 |
| $Q_9$ | -0.0688 | -0.0344 | -0.2562 | -0.1281 |
| $Q_{10}$ | -0.0485 | -0.0343 | -0.0556 | -0.0393 |
| $Q_{11}$ | 0.0835 | 0.0590 | -0.2871 | -0.2030 |
| $Q_{12}$ | -0.0225 | -0.0159 | -0.0565 | -0.0399 |
| $Q_{13}$ | 0 | 0 | 0 | 0 |
| $Q_{14}$ | -0.0284 | -0.0142 | 0.0679 | 0.0339 |
| $Q_{15}$ | -0.0371 | -0.0185 | -0.1763 | -0.0881 |

The electrical field gradient (EFG) generated by the oxygen octahedron on the metal sites M1 an M2 is a tensor value $G_{\alpha\beta}$. The main component $V_{zz}$ of the $G_{\alpha\beta}$ has been calculated and the obtained values are 0.42 and 0.11 e·A$^3$ for Mn1 and Mn2 sites, respectively. Thereby, the Mn1O$_6$ octahedron was found to be ~4 times more distorted than Mn2O$_6$ one. The EFG principal axis lies along the M1 – O1 (2.379 Å) and M2 – O2 (2.088 Å) bonds. The EFG principal axes of Mn2 - Mn1 pair are codirectional and are contradirectional to that of neighbor Mn1-Mn2 pair (Fig. SM1 [22]). This indicates the inversion of the principal axis in the ribbon substructure.

The main results of the structural study on Mn$_2$BO$_4$ may be written as follows: i) the trivalent and divalent Mn ions occupy the M1 and M2 sites, respectively, that leads to the charge ordering in the row Mn$^{2+}$(2)-Mn$^{3+}$(1)–Mn$^{3+}$(1)–Mn$^{2+}$(2); ii) the $Q_2$ and $Q_3$ dominated modes corresponding to the JT distortions are significantly pronounced for Mn1O$_6$ octahedra; iii) strong JT distortion of Mn1O$_6$ octahedra suggests a $d_z^2$ orbital at the Mn1 site; and iv) there is an inversion center of the principal axis of octahedra at the center of the row.

### 3.2. dc Magnetization

Figure 1(a) shows the magnetization curves, as functions of temperature, measured with the applied field perpendicular and parallel to the *c*-axis. It can be seen that the homometallic manganese warwickite is an antiferromagnet with its easy axis of the magnetization perpendicular to the crystalline *c*- axis. Broad maxima near 30 K are pronounced. In the top inset appear the dependences of the derivative *dM*/*dT* as function of the temperature for both directions of the applied field. From these curves we have obtained $T_N$ = 26 K in agreement with the data [12, 20]. In the bottom inset a zoom of the field-cooled (FC) and zero-field-cooled (ZFC) *dc* magnetization curves as a function of temperature for the applied field perpendicular to the *c*-axis are shown. We note that no thermo-irreversibility between the FC and ZFC curves occurs below a critical temperature, which allow suggesting the antiferromagnetic spin arrangement in Mn$_2$BO$_4$.

Figure 1(b) shows the linearization of the inverse *dc* susceptibility (1/$\chi_{dc}$). The Curie-Weiss law is obeyed within an extensive temperature range. The Curie-Weiss temperatures were found to be θ = −118 and -134 K for the applied field parallel and perpendicular to *c*-axis, respectively. The temperatures are negative strongly supporting the antiferromagnetic interactions as dominant. We note, that this result is contrary to the reported data [19], where the positive value of Curie-Weiss temperature has been found. The ratio $\frac{\theta}{T_N} \approx 5$ indicates the presence of disordering components in the magnetic coupling network. The effective magnetic moments per formula unite obtained from Curie-Weiss constants were found being 6.25 and 6.95 $\mu_B$ for the parallel and perpendicular directions of the applied field, respectively. Assuming that both Mn ions are in the high-spin state we calculated the spin component of the effective moment neglecting the orbital component. Accounting for the contribution of each type of Mn ions, the spin component is given by $\mu_s^2 = \sum_i g_i^2 S_i(S_i+1)$. The effective moment per formula unit with one divalent ($S = 5/2$) and one trivalent ($S = 2$) Mn ions and $g = 2$ is $\mu_S = 7.68\ \mu_B$. This value is larger than that experimentally finding. The effective magnetic moment per formula unite taking into account the orbital contribution is expressed by $\mu_J^2 = \sum_i g_i^2 J_i(J_i+1)$. We have estimated the value of $\mu_J$ using the results of work [24], where the orbital magnetism has to be taken into account in the description of electron level energies of real $3d$-ion compounds. For $Mn^{3+}$ ion ($d^4$) $J = 1$ and $g = 1.1$, and for $Mn^{2+}$ ion ($d^5$) it is $J = 5/2$ and $g = 2$. The obtained value is $\mu_J = 6.12\ \mu_B$. The result supposes that the orbital moments of $Mn^{3+}$ ions are not quenched and contribute to the observed magnetic moment. This result also supports the data of neutron diffraction [20], showing the considerably reduced $Mn^{3+}$ magnetic moments $\mu = 1.0\ \mu_B$ relative spin values.

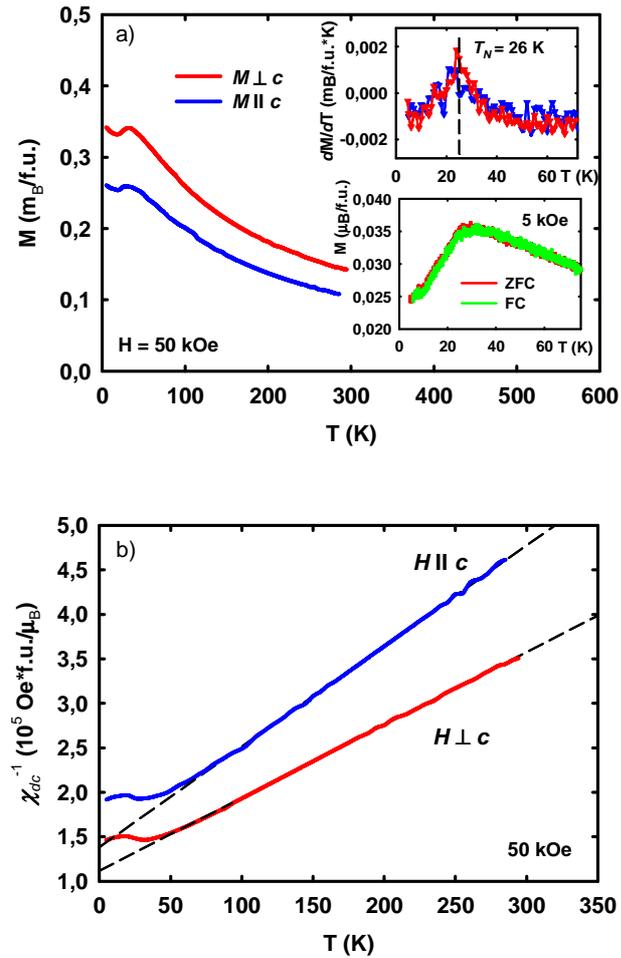

**Fig. 1.** a) The temperature dependencies of *dc* magnetization of $Mn_2BO_4$ single crystal measured in applied field of 50 kOe directed along two orthogonal directions: parallel (H∥) and perpendicular (H⊥) *c*-axis. The top inset shows the peak of the derivative *dM/dT* corresponding to the critical temperature $T_N$. The bottom inset is the field-cooled

and zero-field-cooled magnetization curves measured for the applied field perpendicular to the *c*- axis. b) The inverse susceptibility as a function of the temperature at $H = 50$ kOe.

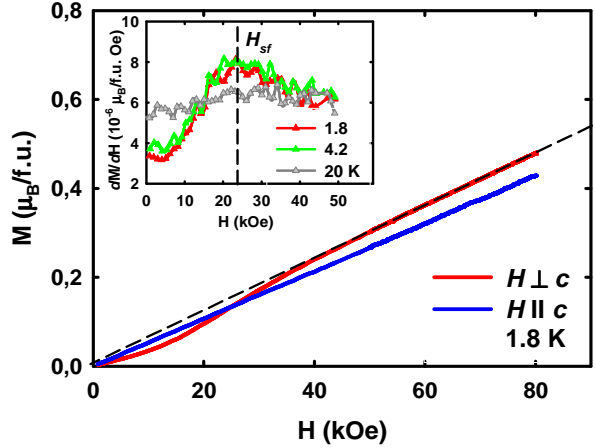

**Fig. 2.** The field dependences of Mn$_2$BO$_4$ magnetization for two orthogonal directions of the applied field, $T = 1.8$ K. The straight line is the extrapolation of the high-field data to the low field range. The spin-flop transition is revealed as the inflection point at $H_{sf}$. Magnetization derivatives in the inset indicate the spin-flop field $H_{sf}$.

Magnetization as a function of applied magnetic field was measured in the temperature range 1.8 – 75 K for both directions of external field. We note that the curves do not display the saturation up to 140 kOe. The isotherms of magnetization at 1.8 K vs applies field are plotted in Fig.2. The magnetization isotherm measured with applied field perpendicular to *c*-axis shows the inflection point at field of $H_{sf} = 24$ kOe, which can be determined as the maxima in the derivative function plotted in the coordinates of *dM*/*dH* vs *H* (top inset). The derivatives of the isotherms measured at 4.2 and 20 K are displayed also. As the temperature increase the curve becomes smooth, nevertheless, the transition can be observed right up to 20 K. We identified the critical field corresponding to the maxima as a spin-flop field.

The magnetization isotherms measured for the applied field parallel to *c*-axis do not show any peculiarities pointing out the easy axis of magnetization lies in the *ab* plane, which confirms the results of the magnetization measurements as temperature function.

### 3.3. Specific heat

Further evidence for AF transition is obtained from specific heat data. The specific heat measurement data at $H = 0$ are shown in Fig. 3(a), where the specific-heat curve is plotted in the coordinates of *C/T* versus *T*. One can see clearly feature at the magnetic ordering temperature $T_N^*$ = 23 K. We note that the specific heat magnitude of Mn$_2$BO$_4$ is higher than those of MgScBO$_4$ in wide temperature range [19], which has not the magnetic contribution to the specific heat. Our attempts to fit the low-temperature data range by the power law $C/T = \gamma + \beta T^2$ have been got the following fitting parameters: are $\gamma = 2.52$ mJ/mol·K$^2$ and $\beta = 9.54$ mJ/mol·K$^4$. The linear term turns out to be of the order of that of the related ludwigites $Co_3BO_5$ and $Co_2FeBO_5$ (3.30 and 3.28 mJ/mol·K$^2$, respectively) [25]. Note that our attempts to measure the transport properties of $Mn_2BO_4$ single crystal have not been successful due to extremely large resistance of the sample (>10$^8$ Ohm at room temperature). Taking into account the insulating nature of all mention materials the question about the origin of this term remains open. The $\beta T^3$ is the contribution due to three-dimensional phonons in the Debye model. The application of this simplified model for the description of the low-temperature data of the specific heat gets too low value of effective Debye temperature than that is expected for an oxyborate rigite structure. So, for correct description critical behavior of the specific heat and for an anomalous part separation, it is necessary to consider other contributions to the specific heat.

The specific heat of a crystalline solid can be expressed by the sum of three main contributions:

$$C_V = C_V^{latt} + C_V^{mag} + C_V^{Sch}, \qquad (1)$$

where $C_V^{latt}$ is the lattice vibration contribution, $C_V^{mag}$ is the magnetic contribution, and $C_V^{Sch}$ is the Schottky contribution. The crystal defects, anharmonic effect, and free electrons contributions are small and can be neglected. The difference between $C_V$ and $C_p$ can be evaluated by the thermodynamic relationship, which requires knowledge of thermal expansion coefficient, material's volume, and isothermal compressibility. However, the difference between $C_V$ and $C_p$ was shown to be important only at high temperatures [26]. In order to quantitatively estimate the critical behavior near $T_N$ the anomalous contribution of the specific heat $\Delta C_V = C_V^{mag} + C_V^{Sch}$ was separated from the measured dependence of the specific heat by subtracting of the regular lattice contribution $C_V^{latt}$. The latter can be expressed using simple model including Debay and Einstein approaches. The lattice contribution to the specific heat given by the equation:

$$C_V^{latt}/R = K_D D\left(\Theta_D/T\right) + K_E E\left(\Theta_E/T\right), \qquad (2)$$

$$D\left(\Theta_D/T\right) = 9 \frac{1}{\left(\Theta_D/T\right)^3} \int_0^{\Theta_D/T} \frac{\left(\Theta_D/T\right)^4 exp\left(\Theta_D/T\right)}{\left[exp\left(\Theta_D/T\right) - 1\right]^2} d\left(\Theta_D/T\right),$$

$$E\left(\Theta_E/T\right) = 3\left(\Theta_E/T\right)^2 \frac{exp\left(\Theta_E/T\right)}{\left[exp\left(\Theta_E/T\right) - 1\right]^2}$$

is shown in the inset fig.3a. Here, $R$ is the gas constant, $\Theta_D$, $\Theta_E$ is characteristic Debay and Einstein temperatures, the $K_D$, $K_E$ is the numerical coefficients. The fit of the temperature range $T > 130$ K gives follow estimations of parameters $\Theta_E = 695$ K, $\Theta_D = 299$ K. The obtained value of $\Theta_D$ is well agreed with that for $MgScBO_4$ ($\Theta_D = 306$ K).

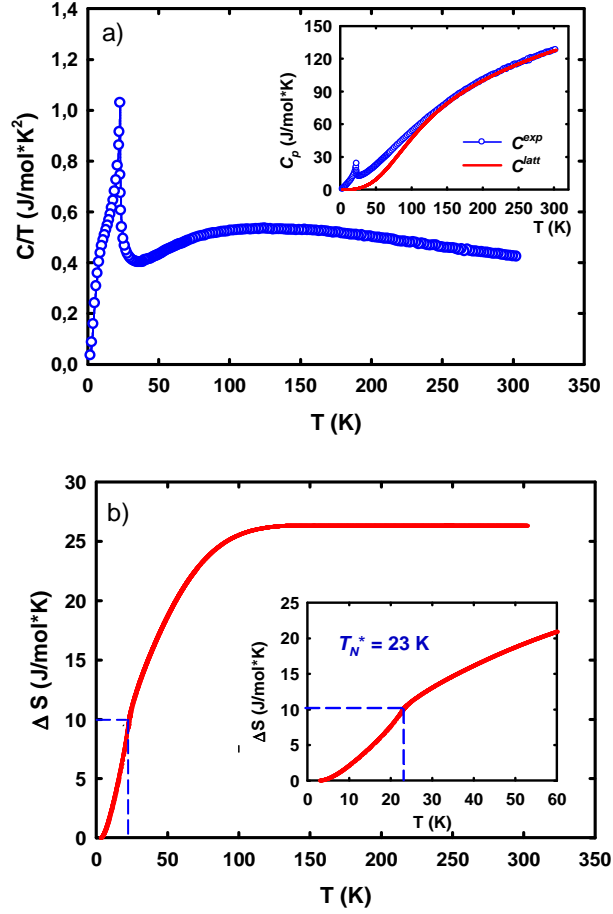

**Fig. 3.** a) Specific heat of $Mn_2BO_4$ plotted as $C/T$ versus $T$. The inset: experimental specific heat (blue circle) and the lattice contribution to the specific heat obtained by fitting to eq. (2) (red solid line). b) Entropy as a function of temperature. The inset: the zoom of the magnetic transition range, $S(T_N) = 10.06$ J/mol·K.

The non-lattice contribution to the entropy ($\Delta S$) including the magnetic ($\Delta S_{mag}$) and Schottky ($\Delta S_{Sch}$) entropies is presented in fig.3b. The full magnetic ordering could accompanied by the entropy released of $\Delta S_{mag} = R[\ln(2S1 + 1) + \ln(2S2 + 1)] = 28.278$ J/mol·K expected for one $Mn^{3+}$ ($S1 = 2$) and one $Mn^{2+}$ ($S2 = 5/2$) ions per formula unite. The entropy at the critical temperature is being $\Delta S(T_N) = 10.06$ J/mol·K, that is less than half of the limit theoretical value of magnetic entropy $\Delta S_{mag}$. Thus, at the temperatures well above $T_N$ the magnetic entropy is associated with the retention of short-range magnetic order in the magnetic spin system. As the temperature increase the magnetic correlations vanish. This fact is consistent with the magnetization data and theoretical calculation of the superexchange interactions.

*4. Uniaxial anisotropy and exchange interactions*

The experimental data have shown that $Mn_2BO_4$ can be considered as standard uniaxial antiferromagnet with easy axis of magnetization lied at the crystallographic *ab* plane. We have estimated the anisotropy constant *K* and anisotropy field $H_a$ values in $Mn_2BO_4$:

$$K(T) = \frac{1}{2}H_{sf}^2(\chi_\perp - \chi_\parallel), \quad H_a = \frac{K}{M_s} \quad (3)$$

where $\chi_\perp$ and $\chi_\parallel$ are the experimental values of magnetic susceptibilities above and below critical field $H_{sf}$, $M_s$ is the sublattice magnetization. Assuming that the average spin per formula unite is $\langle S \rangle = \frac{1}{n}\sum_{i=1}^{n} S_i = 2.25$, where *n* denotes the number of magnetic ions per formula unite, the obtained value is $K(1.8\ K) = 48.03\ erg/cm^3$, which corresponds to $H_a = 165$ Oe. These values are in a good agreement with that reported earlier for $Mn_2B_2O_5$ pyroborate, where only one type of manganese ions presents [27]. The exchange field acting along crystalline *c*-axis was found to be $H_{ex\parallel} = 834$ kOe, while the exchange field inside *ab*-plane is $H_{ex\perp} = 766$ kOe.

In a molecular field approximation for a two-sublattice model with one exchange interaction *J*, the exchange parameter and the Neel temperature are related by simple expression

$$z|J| = \frac{3k_B}{2S(S+1)}T_N, \quad (4)$$

where *z* is the number of nearest neighbors with the exchange interaction *J*, $k_B$ is the Boltzmann constant. Putting the values of $T_N = 26$ K and average spin, we get $z|J| = 5.33$ K.

The consistent approach to analysis of the magnetic behavior of $Mn_2BO_4$ requires calculating the superexchange interaction. We note that the estimations of the superexchange interactions have been made earlier in the work [20], but the interaction through $t_{2g}$-orbits wasn't taken into account, along with the interribbon interactions. We present the deep analysis of superexchange interaction in $Mn_2BO_4$, considering all nonequivalent superexchange pathways of ferromagnetic and antiferromagnetic nature. That allowed to explain observed magnetic behavior and to offer a scenario of the magnetic ordering in $Mn_2BO_4$.

The inter-ion distances inside the ribbon are the order of 3 Å as can be seen from the crystallographic data. Namely, the shortest is $Mn^{3+}$-$Mn^{3+}$ (2.84 Å), and the longest is $Mn^{2+}$-$Mn^{3+}$ (3.36 Å). That is too long for a direct orbital overlapping. The anisotropy energy is much smaller than the exchange one (~0.25 versus ~5 K). We have used the simple model of superexchange interactions [28] applied earlier to the analysis of the complex magnetic structure in $Co_3O_2BO_3$, $Co_2FeO_2BO_3$ ludwigites [1, 5], and $Co_3B_2O_6$ kotoites [30], with a satisfactory agreement with the experimental results. Manganese warwickite $Mn_2BO_4$ belongs to the related family of oxyborate that justifies the analogous consideration. The calculation is restricted by the nearest-neighbor approximation; i.e. only the interactions along the short M-O-M bonds are considered, while the long bonds M-O-M-O-M and M-O-B-O-M are neglected. The ferromagnetic (F) and antiferromagnetic (AF) contributions to Mn-O-Mn couplings exist. Orbitally non-generate $Mn^{2+}$ states ($d^5$) have singly occupied $e_g$ orbitals, that supposes the antiferromagnetic exchange interaction to be dominant. The $Mn^{3+}$ states ($d^4$) have one $e_g$ hole at $d_{x2-y2}$ orbital, that leads to ferromagnetic (F) contributions to the exchange integrals.

The $Mn_2BO_4$ warwickite structure has several types of indirect couplings: 92°, 95°-99°, 102°, and 105°, which can be assigned to 90° exchange interactions, as well as 115°, 119° and 125° exchange interactions. In the 2-1-1-2 row the neighboring cations with common octahedral edges take part in the exchange couplings with bond angles of 99°-105° ($J1$) and 97° ($J2$). The octahedra belonging to the adjacent rows, which are connected by a common edge, allow indirect couplings 92°-105° ($J3$), 96°-102° ($J4$), 95° ($J5$), and 89°-102° ($J6$). The octahedra connected by a common oxygen ion and belonging to the adjacent ribbons allow indirect couplings of 115°, 119°, and 125°, corresponding to $J7$, $J8$, and $J9$ exchange interactions respectively. So, the magnetic structure of $Mn_2BO_4$ can be described by nine exchange integrals $J1$-$J9$, where $J1$-$J6$ are intra-ribbon interactions, while $J7$-$J9$ are inter-ribbon ones. The full set of the orbitals pairs participating in the coupling is presented in Table SM5 [22]. The total integral of cation-cation exchange interaction $J$ can be calculated as a sum of individual orbitals exchange integrals

$$J = \frac{1}{4} \sum_{i,j=1}^{5(d)} \sum_{p=1}^{3} \frac{1}{S_i S_j} I_{ij}^p, \quad (5)$$

where $S_{ij}$ - the interacting cations spins; the sum accounts for the five magnetic ion $d$-orbitals and three $p$-orbitals of the ligand; $I_{ij}^p$ – the superexchange interaction integral between the individual orbitals $i, j$ of two cations via oxygen $p$ orbital. Interactions between two filled or two empty orbitals are neglected.

Taking into account superexchange bonds selected by lattice symmetry and cation distribution we have been wrote the expressions for the exchange integrals as following:

$$J1 = -\frac{1}{20}c\left[\left(\frac{10}{3}b + 2c\right)(U_1 + U_2) - 2bJ_{in}\right] = -4.03 \text{ K};$$
$$J2 = -\frac{1}{16}c\left[\left(\frac{4}{3}b + 2c\right)(U_1 + U_2) - 4bJ_{in}\right] = -1.44 \text{ K};$$
$$J3 = -\frac{1}{16}c\left[\left(\frac{13}{3}b + 2c\right)(U_1 + U_2) - bJ_{in}\right] = -6.83 \text{ K}; \quad (6)$$
$$J4 = -\frac{1}{16}c\left[\left(\frac{10}{3}b + 2c\right)(U_1 + U_2) - 2bJ_{in}\right] = -5.03 \text{ K};$$
$$J5 = J4;$$

$$J6 = -\frac{1}{25}c\left(\frac{16}{3}b + 4c\right)(U_1 + U_2) = -5.21 \text{ K};$$

$$J7 = -\frac{1}{20}\left(\frac{16}{9}b^2 + 2c^2\right)(U_1 + U_2)|\cos 119°| = -2.41 \text{ K};$$

$$J8 = -\frac{1}{20}\left[\left(\frac{4}{9}b^2 + 2c^2\right)(U_1 + U_2) - \frac{4b^2}{3}J_{in}\right]|\cos 115°| = -0.47 \text{ K};$$

$$J9 = -\frac{1}{20}\left[\left(\frac{4}{9}b^2 + 2c^2\right)(U_1 + U_2) - \frac{4b^2}{3}J_{in}\right]|\cos 125°| = -0.65 \text{ K};$$

The factor $|\cos\theta|$ accounts the angle dependence of the transfer parameters. The basic parameters of the model are the ligand-cation excitation energies $U$ ($U_1 = U(Mn^{3+}–O) = 5.0$ eV, $U_2 = U(Mn^{2+}–O) = 4.4$ eV), intra-atomic exchange energy $J_{in}$ ($J_{Mn}^{3+} = 3.0$ eV) and electron transfer parameters $b = 0.02$ (σ-coupling), $c = 0.01$ (π-coupling) defined in the work [29]. The calculated values are presented also. The integrals $J5 = J4$ are equivalent because the similar magnetic ions interact through the same electron orbitals. One can note that the exchange integral values are comparable with those $z|J|$ estimated from the $T_N$, assuming $z = 1$. The intraribbon interactions ($J1$-$J6$) are comparative in magnitude and considerably exceed the interribbon ones. The AF intra-row interaction $J2$ is quenched by ferromagnetic pathway from the $e_g$ hole on $d_{x2-y2}$ orbitals of $Mn^{3+}$ ions. The $J6$ interaction includes antiferromagnetic superexchange pathways only due to singly occupied five $d$-orbitals of $Mn^{2+}$.

To gain insight into the magnetic properties of $Mn_2BO_4$ the division into magnetic sublattices is needed. The number of magnetic sublattices is determined by the different cations number, nonequivalent local cation positions number relative to the principal crystal axes, and interaction sign between the nearest neighbors at last. In $Mn_2BO_4$ the octahedra principal axes have four different directions relative to the cell axes (Fig. SM1). Let warwickite be considered as a magnetic system consisting of eight magnetic sublattices in which crystallographic positions M1 and M2 are divided into four magnetic sublattices: $1a$, $1b$, $1c$, $1d$ and $2a$, $2b$, $2c$, $2d$, that leads to two types of the rows $2a$ - $1a$ - $1b$ - $2b$ and $2c$ - $1c$ - $1d$ - $2d$ (Fig. 4).

The exchange interactions parameters calculated with taking into account the nearest neighbors numbers ($z_{ij}$) are collected in Table 3. The mutual orientation of the sublattices magnetic moments is shown by arrows. The exchange integrals of minimum energy correspond to the strongest coupling. As can be seen the strongest interactions are 2·$J4$ and 2·$J6$ (via two equivalent pathways, $z_{ij} = 2$) between the manganese ions belonging to the chains $1a$-$1c$-$1a$, $1b$-$1d$-$1b$, $2a$-$2c$-$2a$, $2b$-$2d$-$2b$ extending along $c$-axis. The strong inter-chain interaction $|J3| = 6.83$ K favors antiferromagnetic alignment of $Mn^{2+}$ and $Mn^{3+}$ moments. These three interactions reinforce each other and impose the antiferromagnetic alignment between the magnetic moments of $1a$ sublattice and those of $1c$, $2c$ sublattices. The same interactions lead to the magnetic moments of the 1*b* sublattice to be ordered antiferromagnetically with respect to those of 1*d* and 2*d*. This type of the coupling is marked as "ordering interaction" and denoted in bold in Table 3, and with bold lines in Fig. 4, where the calculated local magnetic structure, depicting the short range order, is presented. According to these ordering interactions the arrows directions (↑ or ↓) have been established. The relative weak *J*1, *J*2, *J*8 and *J*9 interactions tend to disturb the AF order imposed by dominant ordering interactions. Such couplings have frustrating charter and are named as "disordering interactions". They are denoted in italic in Table 3 and with red lines in Fig. 4.

In a molecular field approximation for the multisublattice model the exchange field operating on magnetic ions with spin $S_i$, belonging to *i*-th sublattice on the part of other sublattices is given by the expression

$$\boldsymbol{H}_{exi} = \sum_{j=1}^{p} \frac{2J_{ij}}{g_i g_j \mu_B^2} \boldsymbol{\mu}_j, \qquad (7)$$

$$\boldsymbol{\mu}_j = g_j \mu_B \boldsymbol{S}_j.$$

Here $i, j = 1, 2, \ldots, p$ denote sublattice numbers ($p = 8$ at present case), $J_{ij}$ is the exchange interaction parameter between the sublattices, $g_i, g_j$ are spectroscopic factors, $\mu_B$- Bohr magneton, and $S_j$ is spin of the magnetic ion belonging to *j*-th sublattice. In Mn$_2$BO$_4$ the exchange fields acting on the magnetic ions are defined by the competition between ordering and disordering interactions. We have estimated the exchange fields $H_{exi}$ acting on the manganese ions belonging to the 1*a*-1*d* and 2*a*-2*d* sublattices and obtained the values of $H_{ex}^{1a-1d} = 558$ and $H_{ex}^{2a-2d} = 509$ kOe, respectively. Note these values are well agreed with those obtained from *dc* magnetization measurements. The magnetization of each sublattice at the external magnetic field $\boldsymbol{H}$ is described by the Brillouin function $\boldsymbol{M}_i \sim B_{Si}\left(\frac{g_i S_i \mu_B |H + H_{exi}|}{k_B T}\right)$. As the exchange field $\boldsymbol{H}_{exi}$ is lager, the temperature cone of the magnetization $\boldsymbol{M}_i$ is wider and, consequently, the projection of the sublattice magnetization to the select *z*-axis is lower. The fact the exchange field at the site M1 is larger than that at site M2 can arouse the reduction of the magnetic moment at M1 site.

**Table 3.** The exchange interactions integrals (K) in Mn$_2$BO$_4$ warwickite. The strongest ordering interactions are shown in bold. The disordering interactions are shown in italic.

|       | 1a ↑   | 1b ↑   | 1c ↓    | 1d ↓    | 2a ↑   | 2b ↑   | 2c ↓   | 2d ↓   |
|-------|--------|--------|---------|---------|--------|--------|--------|--------|
| 1a ↑  | 0      | *-1.44* | **-10.06** | **-5.03**  | *-4.68* | *-0.47* | **-6.83** | **-2.41** |
| 1b ↑  | *-1.44* | 0      | **-5.03**  | **-10.06** | *-0.47* | *-4.68* | **-2.41** | **-6.83** |
| 1c ↓  | **-10.06** | **-5.03** | 0       | *-1.44* | **-6.83** | **-2.41** | *-4.68* | *-0.47* |
| 1d ↓  | **-5.03**  | **-10.06** | *-1.44* | 0       | **-2.41** | **-6.83** | *-0.47* | *-4.68* |
| 2a ↑  | *-4.68* | *-0.47* | **-6.83**  | **-2.41**  | 0      |        | **-10.42** |        |
| 2b ↑  | *-0.47* | *-4.68* | **-2.41**  | **-6.83**  |        | 0      |        | **-10.42** |

| | | | | | | |
|---|---|---|---|---|---|---|
| 2c ↓ | -6.83 | -2.41 | *-4.68* | *-0.47* | **-10.42** | 0 |
| 2d ↓ | -2.41 | -6.83 | *-0.47* | *-4.68* | **-10.42** | 0 |

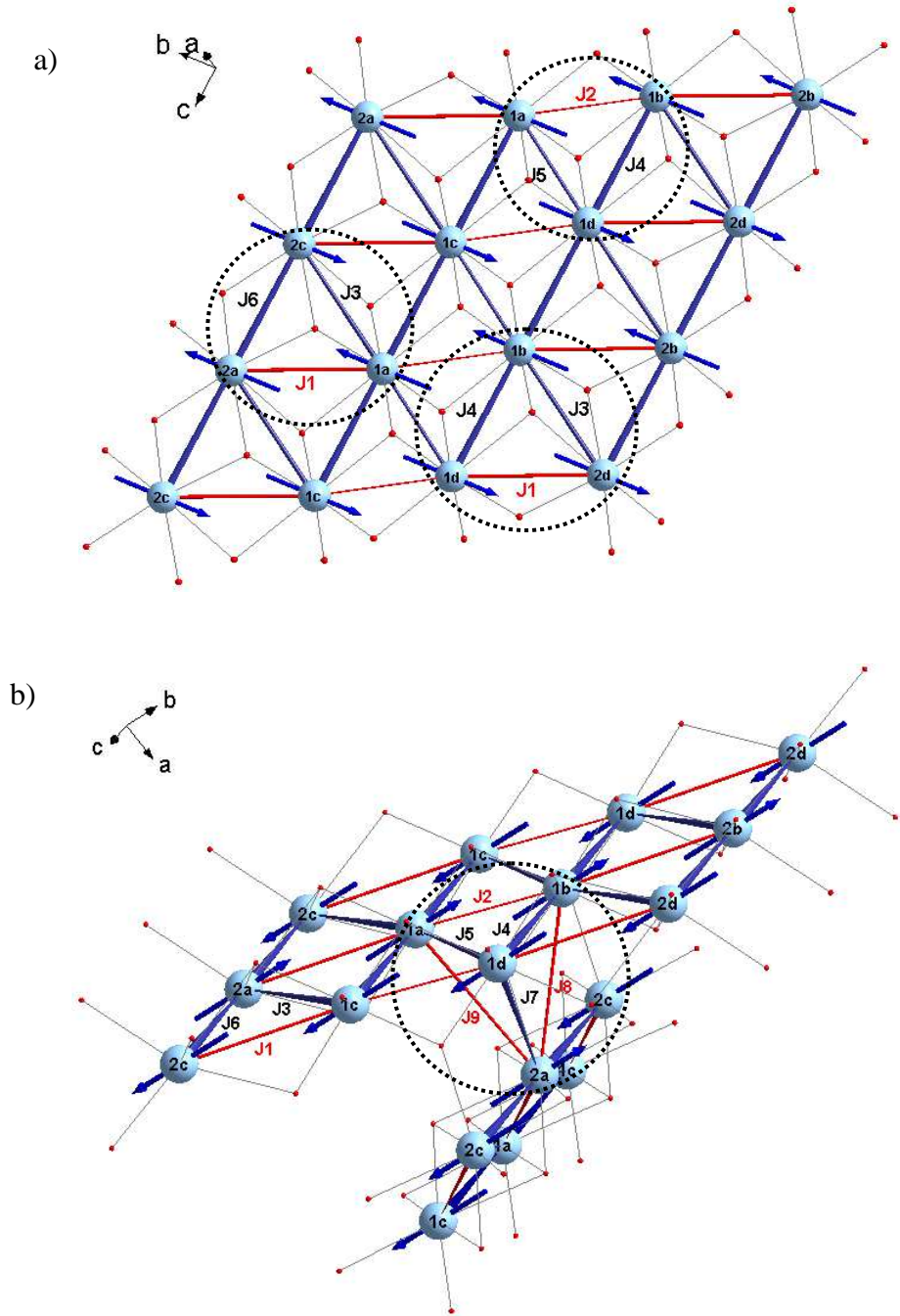

**Fig. 4.** (a) the intra-ribbon indirect exchange interactions (*J*1-*J*6) and b) inter-ribbon ones (*J*7-*J*9) in the $Mn_2BO_4$ warwickite. Numerals indicate the belonging of a crystallographic position to a magnetic sublattice. The frustrated bonds are highlighted red. The interactions strength is shown by the lines thickness. The magnetic moments direction is randomly chosen at the crystallographic *ab*-plane and demonstrates the ordering and disordering bonds. The non-equilateral triangles are highlighted by the circles.

## 5. Discussion

This work presents a careful study of the magnetic properties of homometallic manganese warwickite through *dc* magnetization and specific heat measurements. Magnetic characterization is supported with XRD analysis and superexchange interactions calculations. The X-ray data analysis and normal coordinates calculation have shown that $Mn_2BO_4$ demonstrates the charge ordering of $Mn^{2+}(2)$-$Mn^{3+}(1)$–$Mn^{3+}(1)$–$Mn^{2+}(2)$. The JT distortions give rise to the compression of the $Mn2O_6$ and the considerable elongation of $Mn1O_6$ octahedra along one of the nominal 4-fold axis. The latter suggests a $d_z^2$ orbital at the Mn1 site.

The magnetization measurements have been carried out for two directions of the applied magnetic field: parallel and perpendicular crystalline *c*-axis. A little anisotropy is observed between two directions of the magnetic field. Our magnetic studies has clear revealed the anomaly in the behavior of *dc* magnetization at $T_N = 26$ K which can be unequivocally identified as long-range antiferromagnetic transition. The results of the specific heat strongly support the above-mentioned suggestion, where the $\lambda$-type anomaly was found at $T_N^* = 23$ K, indicating a second-order phase transition. At temperatures $T_N^* < T < 125$ K the short-range magnetic correlations develop, leading to the decrease in the entropy $\Delta S$ and high values of Curie-Weiss temperatures. The easy axis of the magnetization lies in *ab*-plane and spin-flop transition takes place at $H_{sf} = 24$ kOe. The low uniaxial anisotropy $H_a$ is probably associated with orbitally non-degenerate $Mn^{2+}$ state (*S* state). According to the neutron diffraction data the $Mn^{2+}(2)$ moments are being parallel to *b*-axis [20]. One can suppose that the spin-flop transition at $H_{sf}$ is due to the flopping of the magnetic moments of Mn2 - Mn2 inside the antiferromagnetic chain. The depressed values of effective magnetic moments 6.25 and 6.95 $\mu_B$/f.u. obtained for two orthogonal directions of the external magnetic field may be result of the non-quenched orbital moment of $Mn^{3+}$ ions, as well as the intraribbon frustrating superexchange interactions.

The geometric frustrations are underlying by the warwickite structure. There is almost hexagonal arrangement of Mn ions inside the ribbon. The several types of triangular motifs both inside the ribbon and between the adjacent ribbons exist (see Fig. 4 and Table SM3 [22]). Three triangles are resolved inside of the ribbon involving different exchange couplings *J*1-*J*3-*J*6, *J*1-*J*3-*J*4 and *J*2-*J*4-*J*5. The disordering coupling $|J2| = 1.44$ K is weak in compare with $|J4|=|J5| = 5.03$ K and does not break the FM alignment of moments within the row 2-1-1-2. While the rather strong AF coupling $|J1| = 4.03$ K favors an AF alignment of moments $Mn^{2+}$ - $Mn^{3+}$ within the row and leads to considerable frustration of the intra-chains interactions along *c*-axis. The geometry of interribbon bonds such that three types of triangles can be singled out also: *J*4-*J*7-*J*8, *J*2-*J*8-*J*9, and *J*5-*J*7-*J*9. The rather weak interribbon couplings $|J8| = 0.47$ K and $|J9| = 0.65$ K are outweighed by strong $|J7| = 2.41$ K one. So, interribbon interaction *J*7 favors the AF alignment of magnetic moments $Mn^{2+}(2)$ and $Mn^{3+}(1)$ belonging to adjacent ribbons and can be responsible for long-range order in $Mn_2BO_4$. The magnetic frustration experimentally manifests itself in the large ratio of $|\theta|/T_{C,N}$, which is considered as the frustration criteria [31]. For instance, for ferromagnetic materials $|\theta|/T_C \sim 1$, for antiferromagnetic systems, $|\theta|/T_N \sim 2\text{-}5$. For the majority of the warwickites of interest the value $|\theta|/T_{SG}$ is ranged from 8 to 37, which is consistent with a high level of frustration [6, 8, 10]. The value of $|\theta|/T_{SG}$ was found to be ~5 in $Mn_2BO_4$. We conclude, in spite of the magnetic frustration is being in $Mn_2BO_4$ no dramatic frustration level occur, which allow the on-set of long magnetic order at $T_N = 26$ K. In spite of the simplicity of the theoretical method, the calculations have been shown to provide correct description of the magnetic behavior of $Mn_2BO_4$. Therefore, it could be thought of as an available tool for basic and qualitative understanding of magnetic properties of warwickites and related oxyborates.

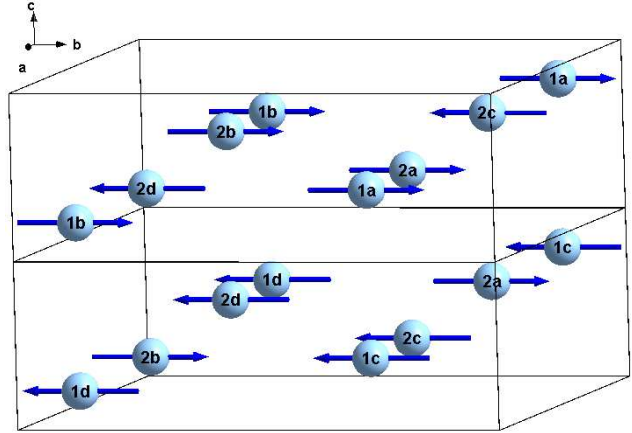

**Fig. 5.** The spin configuration for the ordered state of $Mn_2BO_4$ warwickite resulting from the simple model of superexchange interactions. The double magnetic unit cell along *c*-axis is shown.

In summary, homometallic warwickite $Mn_2BO_4$ can be considered as conventional antiferromagnet showing uniaxial anisotropy. The long-range order occurs below $T_N = 26$ K. The dominant exchange interaction is antiferromagnetic as was found from the superexchange interactions calculations. The strongest interactions are the intra-ribbon ones between the cations along *c*-axis ($J4$ and $J6$). As result, the antiferromagnetic chains 2*a*-2*c*-2*a*, 1*a*-1*c*-1*a*, 1*b*-1*d*-1*b*, and 2*b*-2*d*-2*b* extending along *c*-axis occur. ii) It leads to the doubling of the magnetic cell along the *c*-axis (Fig. 5). We note that a magnetic supercell with twice the volume of the structural cell was found in $Mn_2BO_4$ by neutron diffraction measurements [20], but the doubling direction was not defined. iii) The magnetic coupling between two adjacent ribbons is depressed due to frustrating interactions $J8$, $J9$. The three-dimensional AF ordering is supported by the inter-ribbon interaction $J7$ via the coupling $Mn^{3+}$ - O – $Mn^{2+}$ with the coupling angle 115°, where $Mn^{3+}$ and $Mn^{2+}$ ions belong to the adjacent ribbons.

**Acknowledgments**

The authors wish to thank to Prof. M.V.Gorev for fruitful discussions and the programs Leading Scientific Schools NSch2886.2014.2, Russian Foundation for Basic Research № 12-02-00175-a, 14-02-31051-mol-a, 13-02-00958-a, 13-02-00358-a, and Krasnoyarsk Regional Fund of Science and Technical Activity Support.

**Supporting Information available**: structure parameters, refined bond lengths and angles according to Reitveld refinements of the XRD data, the normal coordinates expressed in Cartesian shifts, the tabulated parameters for superexchange interactions calculation in $Mn_2BO_4$.